\documentclass[fleqn]{article}
\usepackage{amsmath,amssymb}
\usepackage[dvips]{graphicx}

\usepackage{latexsym}

\begin{document}

\pagestyle{plain} 
\setcounter{page}{1}
\setlength{\textheight}{700pt}
\setlength{\topmargin}{-40pt}
\setlength{\headheight}{0pt}
\setlength{\marginparwidth}{-10pt}
\setlength{\textwidth}{20cm}

\title{$p$-th Clustering coefficients $C_{p}$  and Adjacent Matrix for Networks \\
---- Formulation based on String ----   }
\author{Norihito Toyota \and Hokkaido Information University, Ebetsu, Nisinopporo 59-2, Japan \and email :toyota@do-johodai.ac.jp }
\date{}
\maketitle

\begin{abstract}
The phenomenon of six degrees of separation is an old but interesting problem.  
The considerations of the clustering coefficient reflecting triangular structures and its extension to square  one  to  six degrees of separation have been made\cite{Newm21}.  
Recently, Aoyama\cite{Aoyama} has given some considerations to this problem in networks without loops, using a sort of  general formalism, "string formalism". 
In this article, we describe relations between the string formulation proposed by Aoyama and an adjacent matrix. 
Thus we provided a reformulation  of the string formulation proposed by \cite{Aoyama} to analyze networks.  
According to it, we introduced a series of generalized $q$-$th$ clustering coefficients. 
The available rules between  diagrams of graphs and formulae are also given based on the formulation. 
Next we apply the formulation to some subjects in order to mainly check consistency with former studies. 
By evaluating the clustering coefficient for typical networks studied well earlier, we confirm a validity of our formulation.   
Lastly we applied it to the subject of two degrees of separation. 
 
  \end{abstract}
\begin{flushleft}
\textbf{keywords:}
Six Degrees of Separation, String, Clustering Coefficient, Degree Distribution, Generalized Clustering Coefficient
\end{flushleft}

\section{Introduction}\label{intro}

\hspace{5mm} In 1967, Milgram has made a great impact on  the world by advocating the concept 
 "six degrees of separation" by  a celebrated paper \cite{Milg} written based on an social experiment. 
Six degrees of separation  indicates that people have a narrow circle of acquaintances. 
A series of social experiments made by him and his joint researcher\cite{Milg2},\cite{Milg3} suggested 
that all people in USA are connected through about 6 intermediate acquaintances.   
Their studies were strongly inspired by  Pool and Kochen's study \cite{Pool}.    
At the time, however,  numerically detailed studies could not be made 
because computers and important concepts, such as the clustering coefficient needed for a network analysis nowadays, have  not yet developed sufficiently.

One of the most refined models of six degrees of separation was formulated in  work of Watts and Strogatz\cite{Watt1},\cite{Watt2}.  Their framework provided compelling evidence that the small-world phenomenon is pervasive in a range of networks arising in nature and technology, and a fundamental ingredient in the evolution of the World Wide Web. 
Another is the scale free networks proposed by Barabasi et al.\cite{Albe2}, \cite{Albe3}. 
Many empirical networks are characteristic future of scale free \cite{Albe1},\cite{Newm},\cite{Doro1},\cite{Doro2}. 
In spite of furthermore exploring of six degrees of separation\cite{Watt4},\cite{Watt3},  
 they do not examine closely  Milgram's original findings by their model, especially how influence can 
the clustering coefficient proposed in the paper \cite{Watt1} has.   
We have made some study of it in our previous paper \cite{Toyota1} by imposing a homogeneous hypothesis on networks. 
As a result, we found that the clustering coefficient has not any decisive effects on the propagation 
of information on a network and then information easily spread to a lot of people even in networks with 
 relatively large clustering coefficient under the hypothesis; a person only needs dozens of friends.  
Moreover we devoted deep study to the six degrees of separation based on some models proposed 
by  Pool and Kochen \cite{Pool} by using a computer, numerically\cite{Toyota2}. 
 In the article, we  estimated the clustering coefficient along the method  developed by us \cite{Toyota1} 
and improved our analysis of the subject through marrying Pool and Kochen's models to our method introduced in \cite{Toyota1}.   
As a result, it seems to be difficult that six degrees of separation is realized in the models proposed by Pool and Kochen\cite{Pool} on the whole. 

If the network of human relations has a tree structure without loops,  a person connects new persons  in power  of average degree, when he(she) follows his(her)  acquaintances  step by step on his(her) network of human relations. Then the phenomenon of six degrees of separation is not so amazing,  if a person has a few hundred acquaintances. 
A question is that  networks of general human relations include some loop structures. 
This structures decrease the number of new persons that connected with him(her) when he(she) follow  his(her) acquaintances  step by step.  
One of indices characterizing  loop structures is the clustering coefficient. 
Thus it will be important to investigate the effect of the clustering coefficient and degree distribution on the six degrees of separation. 
It is, however, difficult  to investigate the influence of general loop structures.    
There are only  a little research focused on  the effect of loop structures.  

Newman first studied the influence of loop structures in a network  on the subject\cite{Newm21}. 
The study is so stimulating  but only triangle structures and quadrilateral structures on networks were considered.     
It seems to be difficult to  generalize  his framework to $q$-polygon.  
Recently Aoyama proposed the string formulation on the subject\cite{Aoyama}.  
The idea  inspired our study in this article, greatly.  
Unfortunately he considered only tree approximation in the structure of  networks. 
Since he deals with mainly scale free networks,  the approximation is valid up to a point.

In this article we pursue the string formulation and try to discuss the influence of general loops to six degrees of separation. 
One of the aims in this article is to reformulate the string formulation based on an adjacent matrix. 
We can systematically analyze general networks with arbitrary loop structures by the reformulation.  
Next is to check the results derived from it are consistent with results studied so far. 
Then we apply it to the problem of two degrees of separation as a first step so that future prospect 
for the problem are opend.   

 The plan of this article is as follows. 
After introduction,  we reformulate the string formulation by an adjacent matrix in the following section 2. 
First we  explain some notations used in this article and reformulate the string formulation by using an adjacent matrix. 
Here we introduce $R$ matrix that play central role in our formalism.  
According the formalism, we introduce generalized $q$-th clustering coefficients as well as the usual global one.   
Next we give a diagrammatic interpretation for every term appearing in the expansion of the power of $R$ matrix 
like Feynman diagram \cite{Peskin} in quantum field theory.  
In the next section 3, we evaluate some clustering coefficients on some typical networks and discuss the consistency with results investigated by now.  
We shall discuss  the phenomenon of six degrees of separation in the section 4. 
Since any reliable conditions have not been given for $p$-th degrees of separation in networks with strongly connected components  yet,  we can not provide full discussions.  Thus we discuss  two degrees of separation 
and compare the results to those given by Aoyama\cite{Aoyama} in scale free networks.    
We find that the result is a little different from  Aoyama's one\cite{Aoyama}.     
The analysis of an adjacent matrix is seemed to support rather our results. 
We shall devote discussion  on six degree of separation based on the formulation in a subsequent article\cite{Toyota4}.      
The last section 5  is devoted to summary.

\section{String Formulation and Adjacent Matrix}
\label{usage}
We, basically, follow the string formulation introduced by Aoyama \cite{Aoyama}. 
\subsection{Notations}
In this section we describe notations used in this article. 
 We consider a string-like part of a graph with connected $j$ vertices  and call it  "j-string".
$N$ is the number of vertices in a considering network and  $S_j$ is the number of j-string in the network. 
(Note that  $S_j$ in this article is $N$   times larger than $S_{j-1}^{Aoyama}$ defined by Aoyama\cite{Aoyama}.)     
By definition, $S_1=N$ and $S_2$ is the number of edges  in the network. 
$\bar{S}_j$ is the number of nondegenerate  j-string where a nondegenerate string is defined as strings without any multiple edges and/or any loops in the subgraphs as seen in Fig.1. 
We, however, define that the nondegenerate string contains  strings homeomorphic to a circle.  

We call  strings without any loops open string and  strings homeomorphic to a circle closed string. 
Thus we   consider closed strings and open strings in this article. 

It is so difficult to calculate $S_j$ and $\bar{S}_j$, generally. 
Aoyama has calculated  $S_j$ up to $j=7$ but did not supply the explicit in his article\cite{Aoyama}. 
His article says that it would needs dozens of papers if the full expression is described. 
It would be maybe impossible to calculate $S_j$ and $\bar{S}_j$ with $j>7$ explicitly at the present moment.  

 \begin{figure}[t]
\begin{center}
\includegraphics[scale=0.9,clip]{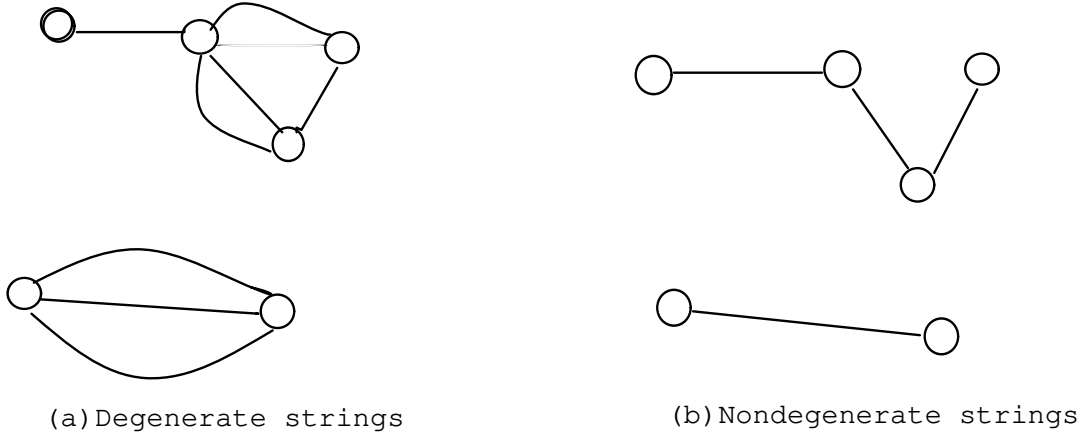} 
\end{center}
\caption{Two types of strings   }
 \end{figure} 
 
\subsection{Clustering Coefficient}
By using the string formulation, we can defined  the usual clustering coefficient which essentially counts the number of triangular structures in a network.   
Let  $\Delta_q$ be the number of polygons with $q$ edges in a network. 
Following Newman \cite{Newm21},  the usual   global clustering coefficient $C_{(3)}$ is  given by 
\mathindent=37mm 
\begin{equation}
C_{(3)}=\frac{6\times \;number \;of \;triangles }{number \;of \;connected \;triplets }=\frac{ 6\Delta_3  }{\bar{S}_3}. 
\end{equation}
We introduce some indices uncovering properties of general polygon structures except for triangle structures in a network. 
From the expression of Eq.(1),  we can generalize it to $q$-$th $ clustering coefficient $C_{(q)}$ straightforwardly;  
\begin{equation}
C_{(q)}=\frac{2p\times \;number \;of \;polygons }{number \;of\; connected \;p\mbox{-}plets }=\frac{ 2p\Delta_p  }{\bar{S}_p}.
\end{equation}

\subsection{Adjacent Matrix Formulation}
We reformulate $C_{(q)}$ introduced in Eq.(2) by utilizing  an adjacent matrix $A=(a_{ij})$.     
Generally the powers, $A^2$,  $A^3$, $A^4$,  $\cdots$ of adjacent matrix $A$ give information as to respecting that a vertex connects other vertices  through $2,3,4, \cdots$ intermediation edges, respectively.   
The information of the connectivity between two vertices in $A^n$ contains multiplicity of edges, generally.    
For resolving the degeneracy, we introduce a new se ries of matrices $R^n$ which give information as to respecting that a vertex connects other vertices  through $n$ intermediation edges without multiplicity. 
We can find it by the following formula\cite{Toyota3}; 
\begin{equation}
R^n=\displaystyle \sum_{i_1,\cdots,i_{n-1}} a_{i_0i_1} a_{i_1i_2}\cdots a_{{i-1},i_{n}} 
\frac{\displaystyle\prod_{i_k,i_j,i_k-i_j>1}^{n}(1-\delta_{i_ki_j})}{(1-\delta_{i_0i_n})}.
\end{equation}
   where the product of $(1-\delta_{i_ki_j})$ of the numerator has the role of protecting of degeneracies from strings and 
  $(1-\delta_{  i_0i_n})$ of the denominator is, however, needed to keep a closed string.

This expression has ($n-1$)ply loops  in a computer program and so it is 
almost impossible to calculate $R^n$ within real time for large $N$ when the rank of $R$ is $N$.   
The expansion of Eq.(3) has $2^{n(n-1)/2}$ terms formally.  
 This value is $32768$ for $n=6$ that is needed in the analysis of six degrees of separation as will be discussed in the later section. 
Though many terms really vanish,  $R^6$ has still so complex expression.  
 We only give the expressions of $R^1\sim R^5$ here and will give that of $R^6$ in Appendix;  
\mathindent=7mm
\begin{align}
[R^2]_{if} &=[A^2]_{if}-[A^2]_{ii}\delta_{if}=[A^2]_{if}-G_{if}, \nonumber \\
[R^3]_{if} &=[A^3]_{if}-\{ G,A \}_{if} +a_{if}, \nonumber \\
[R^4]_{if} &=[A^4]_{if}-\{ G,A^2 \}_{if} +\bigl\{A, diag(A^3)\bigr\}_{if} +2[A^2]_{if}  +[G^2-G-AGA]_{if}+3a_{if}[A^2]_{if} \nonumber \\
[R^5]_{if} &=[A^5]_{if} -\bigl\{A, diag(A^4)\bigr\}_{if}-\{ G,A^3 \}_{if} -\bigl\{A^2, diag(A^3)\bigr\}_{if}
+3\bigl([A^2]_{if}\bigr)^2 [A]_{if} \nonumber  \\
&+3[A^3]_{if}[A]_{if}+2\{ G^2,A \}_{if}+[GAG]_{if}- 6\{ G,A \}_{if} -\{ AGA,A \}_{if} +3[A^3]_{if}\nonumber \\ 
&   +\bigl\{A, diag(AGA)\bigr\}_{if}+2[diag(A^3G)]_{if} -[A\cdot diag(A^3)\cdot A]_{if} \nonumber - [diag(A^3)]_{if} \nonumber \\
&+3\sum_{k} a_{ik}a_{kf} \Bigl( [A^2]_{kf} + [A^2]_{ik}-\delta_{if}[A^2]_{kf} \Bigr) +4a_{if}\bigl( 1-a_{if} \bigr), 
\end{align}
where suffix is abbreviate in trivial cases and $\{\cdot, \cdot\}$ means the anticommutation relation;  $\{A,B\}=AB-BA$. 
$diagA$   indicates the diagonal matrix  whose elements are the diagonal elements of $A$,  and 
$G$ is the diagonal matrix defined by
\mathindent=37mm 
\begin{eqnarray}
G&=&\left[
\begin{array}{cccc}
 k_1&0&0&\cdots \\
 0&k_2&0 &\cdots\\
 0&0&k_3 &\cdots\\
  \vdots &\vdots&\vdots&\ddots \\
   \end{array}
   \right] , 
\end{eqnarray}
where $k_i$ is the degree of vertex $i$. 

By using $R^n$, $\bar{S}_p$ and  generalized $p$-th clustering coefficient $C_{(p)}$ are given by
 \begin{equation}\bar{S}_p=\sum_{i,j} (R^{p-1})_{ij}/2,\end{equation}
\begin{equation} 
C_{(p)}=\frac{\mbox{Tr} R^p  }{ \displaystyle \sum_{i,j}R^{p-1}}, 
\end{equation}
where denominator and numerator indicates the contribution from open strings and a closed string, respectively.  
Thus usual clustering coefficient $C_{(3)}$ becomes 
\begin{equation}\displaystyle 
C_{(3)}=\frac{\mbox{Tr} R^3  }{ \displaystyle \sum_{i,j} (A^2)_{ij}-(A^2)_{ij} \delta_{ij}  }=\frac{ \mbox{Tr} A^3  }{ 
 ||A||-\mbox{Tr} A^2 }.
\end{equation}
where we introduced a new symbol $|| \cdots ||$ which denotes   $ ||A|| \equiv \sum_{i,j}A_{ij} $.

\subsection{Diagrammatic Interpretation}
The expression of $R^n$ is rapidly  complicated as $n$ increases.  
We, however, notice that every term appearing in the expansion of $R^n$ closely corresponds 
to a certain graph\cite{Toyota3}.    
So if a certain graph is given, 
we can write out the expression corresponding to it like Feynman's rule in quantum field theory\cite{Peskin}.   
We describe the diagrammatic construction of $R^n$ based on the relation between both.
 
\begin{enumerate}
\item
Draw all graphs with $n$ edges including degenerate graphs except for closed string. 
\item
Assign sign $(-1)^{n-1+v}$ for every graph where $v$ is the number of vertices included in the graph.  
\item
Calculate degeneracy index $m$ defined in the following Eq.(9) and it is the coefficient of the term corresponding to
the graph; 
   \begin{equation}
m=\displaystyle \prod_{i,k_i \neq 1} [ \frac{k_i-1}{ 2} ]_{Ga},
\end{equation}
where $[\;\; ]_{Ga}$ means Gauss symbol.  
The coefficient supply essentially the number of Euler paths  in a graph. 
We do not distinguish a path and its opposite path and 
start from vertex with hight degree in the cases that there is not any odd vertex in the graph.   
\end{enumerate}

Thus the subject of finding the expression of $R^n$ reduces to the estimation of   
diversity of graph with certain constant number of edges and the number of Euler paths in the graph.   

We describe the typical relations between  expressions and graphs. 
The typical graphs are given in Fig.2 and Fig.3 where the arrows indicate the order in Euler paths 
and  i and f shows an initial  vertex and a final one, respectively. 
These graphs correspond to the following expressions;
   \begin{eqnarray}
Figure 2 (a) &=& k_i,\\
Figure 2 (b) &=& a_{if},\\
Figure 2 (c)  &=& (A^3)_{ii},\\
Figure 3 (a)  &=& (A^2)_{if}, \\
Figure 3 (b)  &=& a_{ij}(A^2)_{if}.
\end{eqnarray}
It is assumed that the right vertex has not any other edges in (a) of Fig 2. 
The generalization of Eq.(12) to polygons with $n$ vertices is easily given by  $(A^n)_{ii}$.  
Eq.(13) is also straightly extended to $(A^n)_{if}$ in the cases that $n$ vertices are linearly  connected. 

Degenerate multigraphs also reduce multigraphs with little multiplicity as shown in  Fig.4. 
The left graph comes down to the  corresponding expression $a_{if}$ in (a) of Fig.4.  
Since these are only  useful  relations between graphs and expressions, 
these graphs should be independently considered in the step 1 in the diagrammatic construction. 
These  correspondence may be rather effective in estimating explicit expression of $R^6$.    

\begin{figure}[t]
\begin{center}
\includegraphics[scale=0.6,clip]{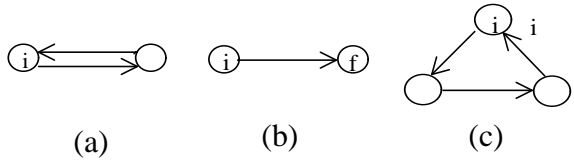} 
\end{center}
\caption{Typical diagrams　I }
 \end{figure} 
 
 \begin{figure}[t]
\begin{center}
\includegraphics[scale=0.6,clip]{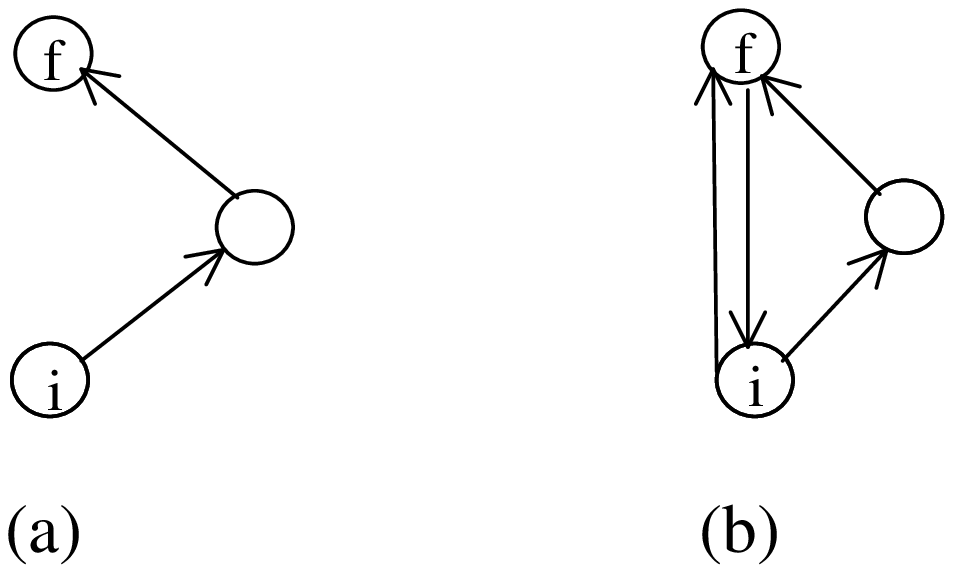} 
\end{center}
\caption{Typical diagrams II}
 \end{figure} 
 
\begin{figure}[t]
\begin{center}
\includegraphics[scale=0.6,clip]{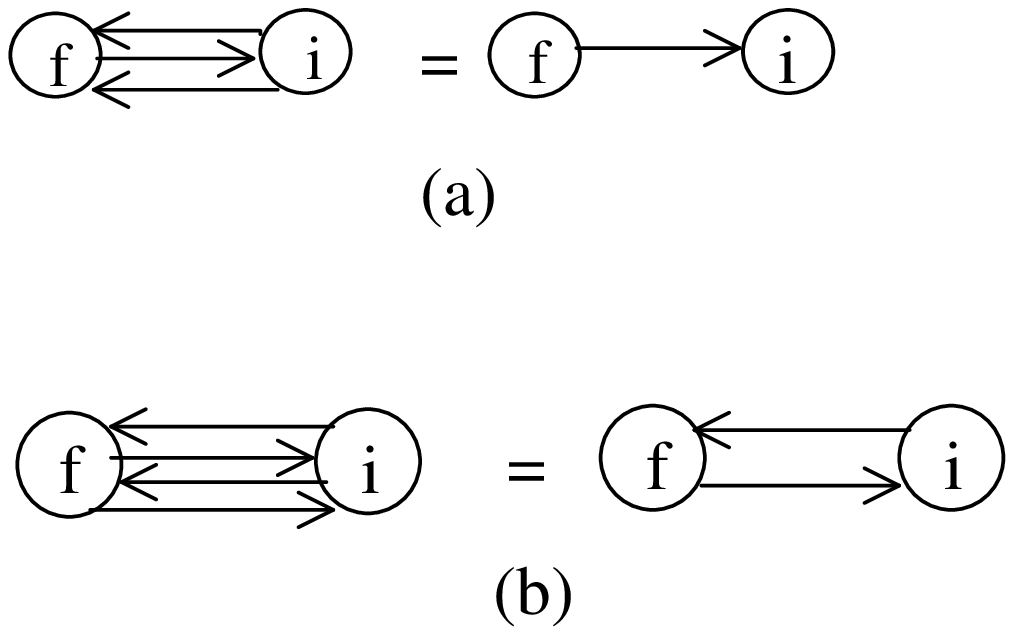} 
\end{center}
\caption{Reducing of diagrams }
 \end{figure} 

\subsection{An Example}
We give an example of the diagrammatic construction of $R^n$. 
As a nontrivial case, we consider the construction of $R^4$ which is constructed from graphs with four edges. 
All topologically independent graphs with four edges are given in Fig.5. 
  For every graph, signs and coefficients derived from step 2 and 3  in the diagrammatic construction 
are as follows; 
   \begin{eqnarray}
Figure 5 (a) &=& [A^4]_{if},\\
Figure 5 (b) &=& [G^2]_{if},\\
Figure 5 (c)  &=& -\{A^2,G\}_{ii},\\
Figure 5 (d)  &=& -[G]_{if}, \\
Figure 5 (e)  &=& -[AGA]_{ij},\\
Figure 5 (f)  &=& 2[A^2]_{if},\\
Figure 5 (g)  &=& A^3_{ii}a_{if}+ A^3_{ff}a_{if},  \\
Figure 5 (h)  &=& 3a_{ij}[A^2]_{if}.
\end{eqnarray}

It is easy to confirm that the sum from Eq.(15) to Eq.(22) give $R^4$ in Eq.(4).  

\begin{figure}[t]
\begin{center}
\includegraphics[scale=0.5,clip]{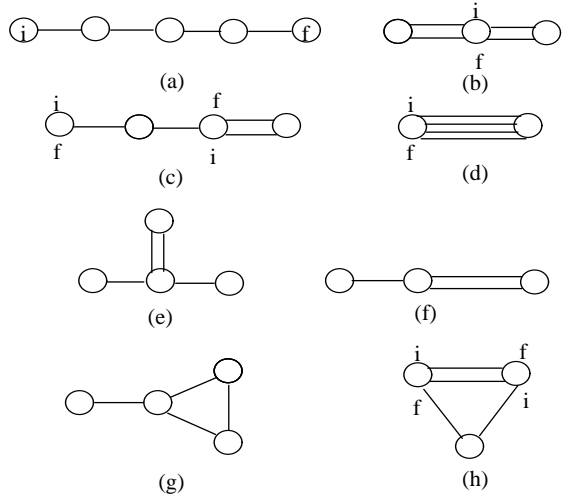} 
\end{center}
\caption{Topologically independent graphs with four edges }
 \end{figure}

\section{Clustering Coefficient of Some Networks}
In this section we calculate $C_{(3)}$ in some typical networks by using the formulation given 
in the previous section. 
Thus we investigate consistency of our formulation with results observed so far. 
We adopt the configuration model \cite{Bebe}.\cite{Bend},\cite{Moll} in producing networks. 
The model can systematically produce networks with arbitrary degree distribution. 
But the networks produced by the model are degenerate multigraphs, generally. 
We modify it a little to produce networks without multiplicity. 
Since it is not essential in this article, we omit the technical details of it. 

First we study random networks\cite{Erdos1} where the degree distribution is Poisson distribution\cite{Albe2}. 
The $N$ dependence on $C_{(3)}$ given by  computer simulations is shown in Fig.6. 
Theoretically it is shown that the clustering coefficient behaves as $C_{(3)}=\langle k \rangle/N$ in random networks where $\langle k \rangle$ indicates average degree \cite{Albe1}.  The log-log plot in Fig.6 shows linear behavior with slope nearly 1. 
Thus   $N$ dependence of $C_{(3)}$  is consistent with observations so far.

Moreover we can observe similar scaling law in networks with the normal distribution in degree distribution.   
As Fig.7,  all the slopes of the log-log plot indicating $N$-$C_{(3)}$ relation are also nearly 1 for various values of $\langle k \rangle$ and the standard deviation $\sigma$.  
The delicate difference of the values of the slopes was discussed in \cite{Toyota3}.   
 
  \begin{figure}[t]
\begin{center}
 \includegraphics[scale=0.7,clip]{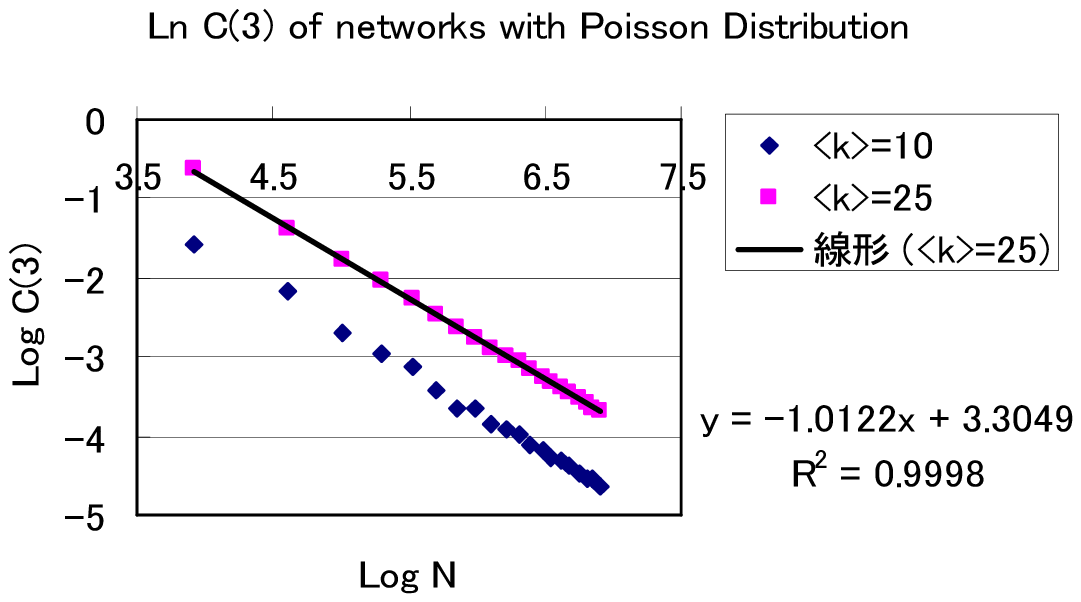}  
\end{center}
\caption{The scaling of $C_{(3)}$ in random graphs with Poisson distribution}
 \end{figure} 

\begin{figure}[t]
\begin{center}
 \includegraphics[scale=0.7,clip]{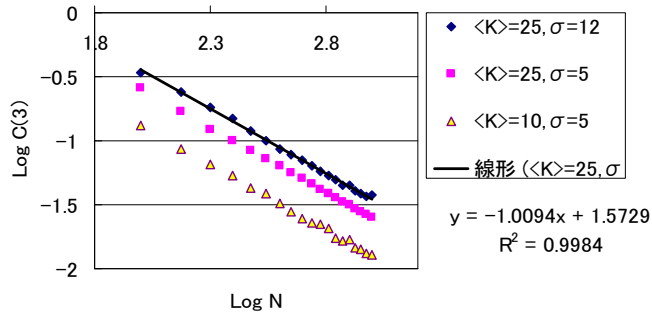}  
\end{center}
\caption{The scaling of $C_{(3)}$ in graphs with Normal distribution}
 \end{figure} 
 
 \begin{figure}[b]
\begin{center}
 \includegraphics[scale=0.35,width=8cm,height=4.5cm,clip]{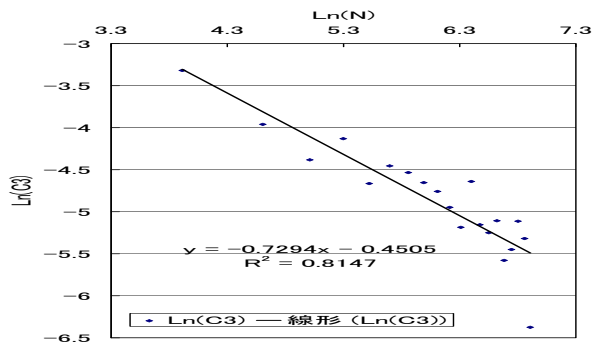}  
\end{center}
\caption{The scaling of $C_{(3)}$ in scale free networks with $\langle k \rangle=4$}
 \end{figure} 
 
It is known that the clustering coefficient follows the scaling law $C_{(3)} \sim N^{-0.75}$ for scale free networks\cite{Albe1} with $\langle k \rangle=4$.  
Our simulation results for scale free networks that have the average degree 4 are given by 
Fig. 8. 
The slope, about -0.73, of the approximative line in Fig.8 shows that the above result almost agrees with  
current one. 
This also  strengthens the validity of presented formulation.

\section{Application to Two Degrees of Separation}
Aoyama has proposed a condition, so-called Milgram Condition, for $p$-$th$ degrees of separation\cite{Aoyama}; 
\begin{equation}
M_p \equiv \frac{\bar{S}_p}{N} \sim O(N).
\end{equation}
For six degrees of separation, we obtain from Eq.(6)
\begin{equation}
\bar{S}_7= \sum_{i,j} (R^6)_{ij}/2. 
\end{equation}



Before evaluating $R^6$ for six degrees of separation, we as a first step study 
two degrees of separation to check consistency in this article. 
Since 2-string cannot have any loops,  we have only to consider graphs by the tree approximation in the study of two degrees of separation.  

Fig. 9 and Fig.10  show $M_2$  and $N$  in Eq.(23) for random networks  
with Poisson distribution  and networks with the normal distribution in degree distribution. 
From the figures, we observe a relation $M_2 \sim const.$, independently of $N$.  
The linear lines show $N$ in the right hand side of Eq.(23). 
Though one intersection  point in the  Fig.9 and Fig.10 only makes Milgram Condition,   
$M_2>N$ means that two degrees of separation is realized. 
It is getting quite difficult that the Milgram Condition is satisfied for small $\langle k \rangle$. 
Though $N \leq1000$ in our simulations,  we can also speculate where is the intersection point between two lines  for large $\langle k \rangle$ 
due to  $M_2 \sim const.$ for Normal distribution.
Moreover we notice the constant values are almost proportional to $\langle k \rangle^2$.  
It is natural that such situation occurs because loop structures can be neglected in two degrees of separation. 
 Thus we can approximately estimate the critical point that the Milgram Condition is satisfied  for arbitrary $\langle k \rangle$.  
 In this connection,  we obtain  $\langle k \rangle \sim (\mbox{a few}\times 10^4)$ for $N\sim (\mbox{a few}\times 10^8)$ which is about the population of USA, in a random network.  
We think that the value of $\langle k \rangle$ is   as large as it is unrealistic, because some social researchers estimated the average number of  acquaintances of a person is 290 \cite{Bernard1},\cite{Bernard2},\cite{Bernard3}. 

\begin{figure}[h]
\begin{center}
\includegraphics[scale=0.9,width=8cm,height=5cm,clip]{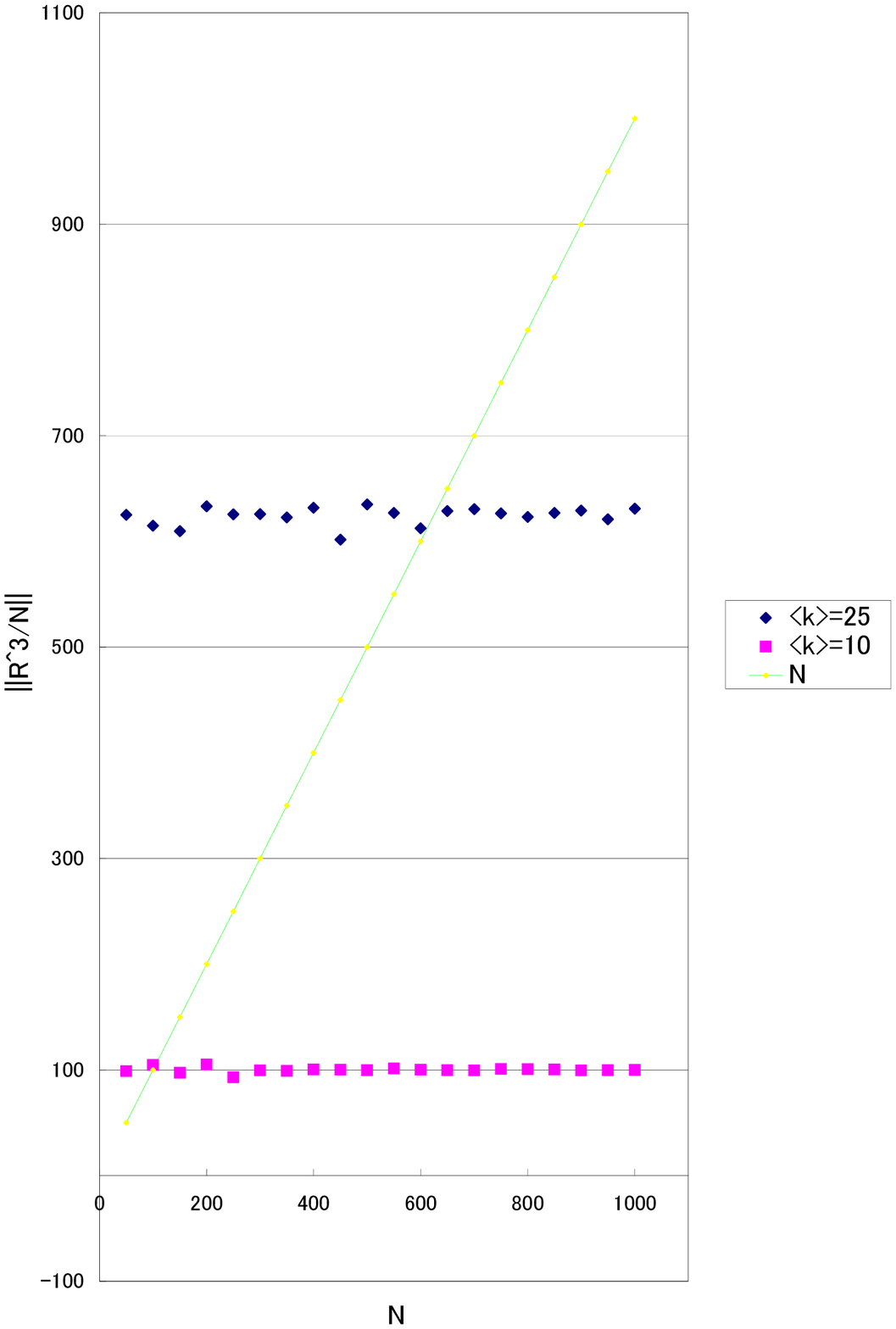}
\caption{$M_2$  with $\langle k \rangle=10$ and $\langle k \rangle=20$ for Poisson  distribution   }
\includegraphics[scale=0.9,width=9cm,height=5cm,clip]{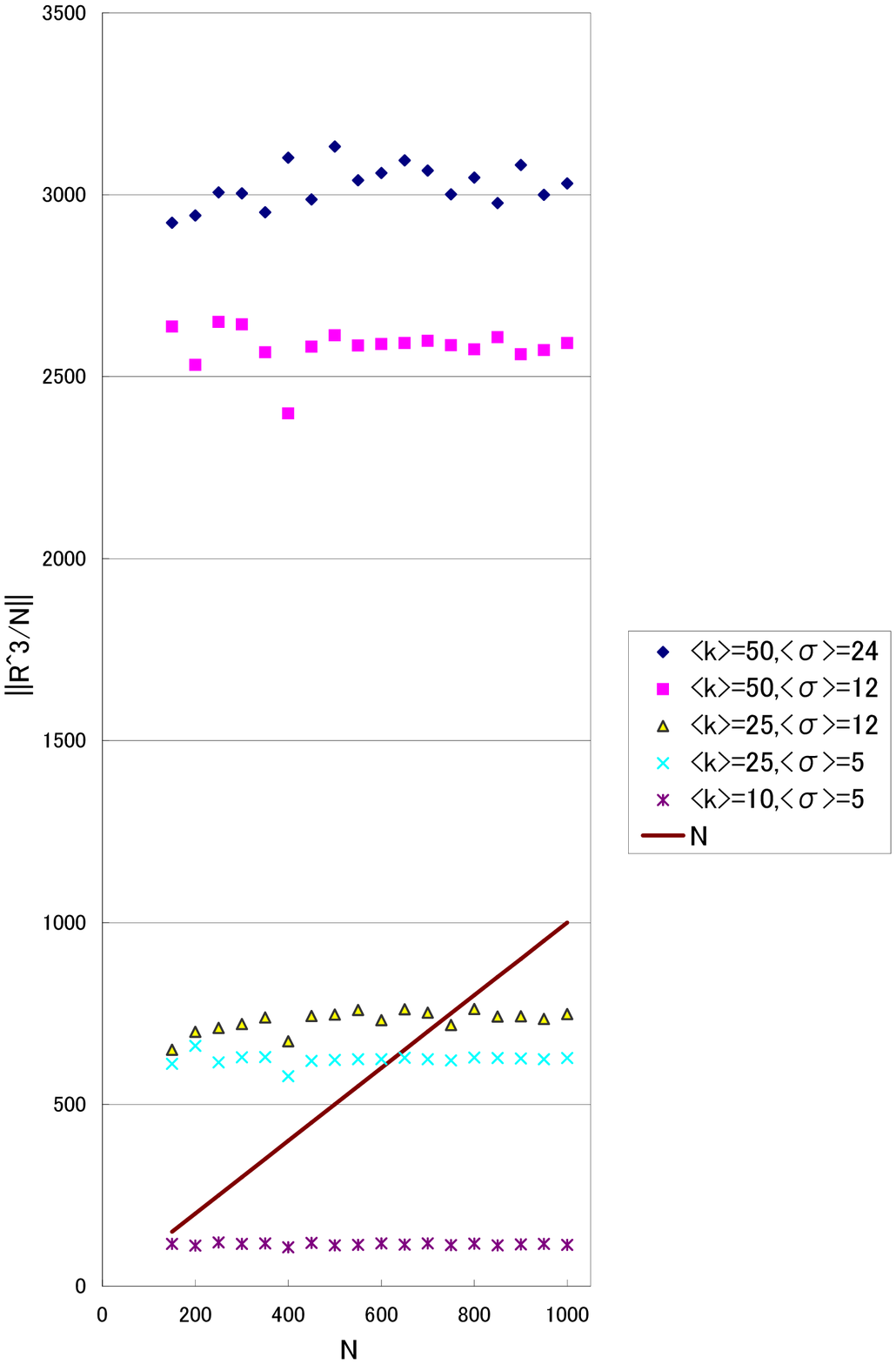}
\caption{$M_2$ with $(\langle k \rangle,\sigma) =(10.5),(25.5),(25,12),(50.12), $
$(50,24)$ for Normal distribution } 
\end{center}
 \end{figure} 
 

\begin{figure}[t]
\begin{center}
\includegraphics[scale=0.8,width=10cm,height=5cm,clip]{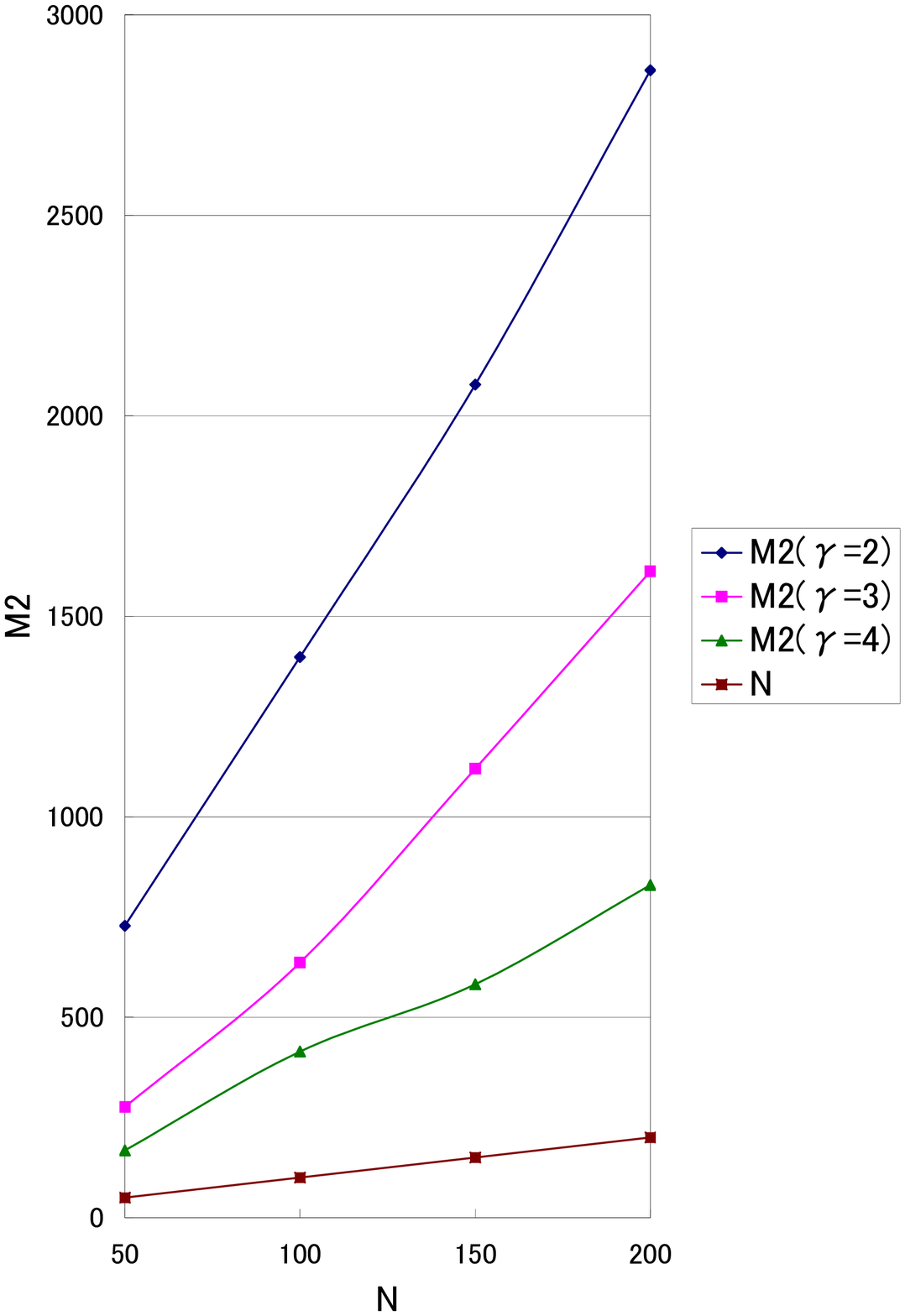 }
\caption{$M_2$ and $N$ for $\gamma=2,3,4$ in SF networks}
\end{center}\end{figure}
\vspace{6mm}

\begin{figure}[t]
\begin{center}
\hspace{3cm} 
\includegraphics[scale=1.7,width=11cm,height=7cm,clip]{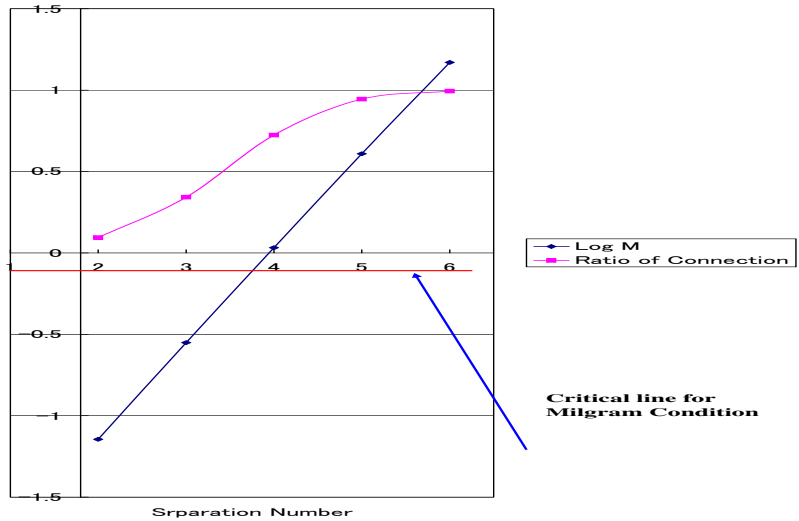}
\caption{$M_2/N$ and Connection Ratio v.s. degree number of separation for SF network with $\gamma=2$ }
\end{center}
 \end{figure} 

The behavior of $M_2$ is rather different from the Normal and Poisson random networks for scale free networks with degree distribution $P(k) \sim k^{-\gamma}$. 
$\langle k \rangle$  depends on the index $\gamma$ in the scale free networks  produced by the configuration model.  	
As $N$ increases, so $M_2$ increases quickly than $N$ as is shown in Fig.11. 
The increasing rate is larger for the smaller index . 
Aoyama\cite{Aoyama} has pointed out that  critical index is $\gamma=2$  in two degrees of separation.
Fig..11, however, shows that $\gamma\leq 2$ can not realize the two degrees of separation.   
The result is different from Aoyama's one. 

For $\gamma=2$,  $M_p/N$ and the ratio of the zero components in $A^2$ in our formalism are given in  Fig.12 where $p$ means the $p-$th degrees of separation ("separation number").     
Evaluating the ratio of the zero components, $\gamma\leq 2$ can not realize  two degrees of separation but
rather four degrees of separation  is realized where 75 percent of all the nodes are connected.  
From (23), the critical velue of $\log_{10} M_2/N$ would be an upper areas of the line a little smaller than zero.  
The  $M_p/N$ satisfying the Milgram condition in Fig.12 also support this results. 
So, our formalism is thought to be available  in order to analyze degree numbers of separation. 
   We shall devote discussion  on six degree of separation based on the formulation in a subsequent article\cite{Toyota4}. 
      
\section{Summary}
In this article, we provided a reformulation of the string formulation proposed by \cite{Aoyama} to analyze networks.  
Fusing adjacent matrix into the formalism, we reformulate the string formalism.  
According to it, we introduced a series of generalized $q$-$th$ clustering coefficients. 
Their function is not yet considered in this article and left as a theme for research in the future. 
Then we introduce the  $R$ matrix in the formalism developed in this article instead of $A$.  
The power of $R$ plays central role in the analysis of this article.    
Every term of the expansion of $R^n$  can be also interpreted graphically and it would make a projection for the future
  in the estimation of $R^n$ that has complex expressions for large $n$.  

On the latter half of this article, we apply the formulation to some subjects in order to mainly check consistency with former studies. 
We first evaluated the clustering coefficient for typical networks studied well earlier. 
We could  confirm a validity of our formulation by these in some degree.  
Lastly we applied the formulation to the subject of two degrees of separation. 
We find that the result is a little different from Aoyama's one\cite{Aoyama}; the separation number is not two but four at $\gamma=2$.    
It is noticed that by analyzing the number of zero-components in $A^2$, our results are rather supported. 

The following problems are yet left in future:
\begin{enumerate}
\item
Finding  explicit  expressions of $R^n$ for arbitrary $n$ by applying our formalism, especially diagrammatic construction.  Then finding  a general formula for arbitrary $n$ from the series of expression. 
  \item
Revealing relations between $p$-$th$ degrees of separation and $N$, $\langle k \rangle$ or $<k^n>$.    
 More definitely,  discovering the relations between $p$ and  $N$, $\langle k \rangle$ or $<k^n>$. 
   \item
 Revealing relations between $p$-$th$ degrees separation and various loop structures, especially $C_{(q)}$.
\end{enumerate}
The last one will be discussed in the subsequent paper \cite{Toyota4}.   
This article give a sort of general formalism to investigate above problems, including  preliminary studies of them.



\newpage

\appendix
\section{$R^6$ and Tr$R^n$}
In this appendix we give an explicit expression $R^6$ and the expressions of Tr of $R^n$ for $n=1\sim6$.  
 $R^6$ is obtained after straightforward but long tedious calculation. 
We divide it into the following four parts to   brighten the prospects of the caluculation. 
\mathindent=0mm


\begin{align}
[R^6]_{if} =&\sum_{j,k,l,m,n} a_{ij} a_{jk}  a_{kl} a_{lm}  a_{mn} a_{nf} \times 
\Delta_{ik}  \Delta_{jl} \Delta_{km}  \Delta_{ln} \Delta_{mf}  \Delta_{il} \Delta_{jm}  \Delta_{kn}\Delta_{lf}  \Delta_{im} \Delta_{jn}  \Delta_{kf}\Delta_{in} \Delta_{jf} \nonumber \\
=&\sum_{j,k,l,m,n} a_{ij} a_{jk}  a_{kl} a_{lm}  a_{mn} a_{nf} \times 
 \Delta_{ik}  \Delta_{jl} \Delta_{km}  \Delta_{ln} \Delta_{mf}  \Delta_{il} \Delta_{jm}  \Delta_{kn}\Delta_{lf}  \Delta_{im} \Delta_{jn}  \Delta_{kf} \nonumber \\
&-\sum_{k,l,m,n} a_{if} a_{fk}  a_{kl} a_{lm}  a_{mn} a_{nf} \times \Delta_{ik}  \Delta_{fl} \Delta_{km}  \Delta_{ln} \Delta_{mf} \Delta_{il}  \Delta_{kn}  \Delta_{im} \nonumber \\
&-\sum_{j,k,l,m} a_{ij} a_{jk}  a_{kl} a_{lm}  a_{mi} a_{if} \times \Delta_{ik}  \Delta_{jl} \Delta_{km}  \Delta_{li} \Delta_{mf}  \Delta_{jm} \Delta_{lf} \Delta_{kf}  \nonumber \\
&+\sum_{k,l,m} a_{if} a_{fk}  a_{kl} a_{lm}  a_{mi} \Delta_{ik}  \Delta_{fl} \Delta_{km}  \Delta_{li} \Delta_{mf},   \nonumber \\
\equiv & R^6[1]_{if} +  R^6[2]_{if}+R^6[3]_{if}+  R^6[4]_{if},
\end{align}
where $\Delta_{ik} = 1-\delta_{ik}$. 
Furthemore we divide $R^6[1]_{if}$ into the following four parts to   brighten the prospects of the caluculation. 

\begin{align}
 R^6[1]_{if} =&\sum_{j,k,l,m,n} a_{ij} a_{jk}  a_{kl} a_{lm}  a_{mn} a_{nf} \times 
\Delta_{ik}  \Delta_{jl}  \Delta_{km}   \Delta_{ln} \Delta_{mf}  \Delta_{il} \Delta_{jm}  \Delta_{kn}\Delta_{lf}  \Delta_{im} \Delta_{jn}  \Delta_{kf} \nonumber \\
 =&\sum_{j,k,l,m,n} a_{ij} a_{jk}  a_{kl} a_{lm}  a_{mn} a_{nf} \times  \Delta_{ik}  \Delta_{jl} \Delta_{km}  \Delta_{ln} \Delta_{mf}  \Delta_{il} \Delta_{jm}  \Delta_{kn} \Delta_{lf}   \Delta_{jn} \nonumber \\
-& \sum_{j,k,l,n} a_{ij} a_{jk}  a_{kl} a_{li}  a_{in} a_{nf} \times \Delta_{ik}  \Delta_{jl} \Delta_{ln} \Delta_{if} \Delta_{kn}  \Delta_{lf} \Delta_{jn} \Delta_{kf}  \nonumber \\
-& \sum_{j,l,m,n} a_{ij} a_{jf}  a_{fl} a_{lm}  a_{mn} a_{nf} \times \Delta_{if}  \Delta_{jl} \Delta_{fm}  \Delta_{ln} \Delta_{il}  \Delta_{jm}  \Delta_{jn} \nonumber \\
+& \sum_{j,l,n} a_{ij} a_{jf}  a_{fl} a_{li}  a_{in}  a_{nf} \Delta_{if}  \Delta_{jl} \Delta_{ln}  \Delta_{jn} \Delta_{mf},   \nonumber \\
 \equiv & R^6[1,1]_{if} +  R^6[1,2]_{if}+R^6[1,3]_{if}+  R^6[1,4]_{if}. 
\end{align}

The four terms are respectively expressed as follows;

\begin{align}
 R^6[1,1]_{if} =&[A^6]_{if}+[A^4]_{if} \bigl(4-(k_i+k_f) \bigr) 
+[AGA]_{if}(k_i+k_p)-\{AGA,A^2 \}_{if}-[A^2GA^2]_{if} 
 \nonumber \\
 &+2[A(G^2-3G)A]_{if}  +3\sum_{j,k} a_{ij}a_{jk}a_{kf}[A^2]_{jk}-\sum_j [A^3]_{jj} 
   \bigl(  a_{ij}[A^2]_{jp}+ [A^2]_{ij} a_{jf}\bigr) \nonumber \\
+&2\sum_j [A^2]_{ij} [A^2]_{jf} \bigl(  a_{ij}+ a_{jf}\bigr) 
+ [A^2]_{if} \bigl( k^2_i+k^2_f -3(k_i+k_f)+4\bigr) \nonumber \\
- &[A^3]_{if}\bigl(  [A^3]_{ii}+ [A^3]_{ff} \bigr) + \bigl([A^3]_{if} \bigr)^2 
  +  \sum_{j} a_{ij}a_{jf}\Bigl( \bigl( [A^3]_{ij} + [A^3]_{fj}\bigr)\nonumber \\
 -& [A^4]_{jj} -2\bigl( [A^2]_{ij} + [A^2]_{fj}\bigr)+ [AGA]_{jj}  
+ \bigl( ([A^2]_{ij})^2 + ([A^2]_{fj} )^2 \bigr) \Bigr) \nonumber \\
+&\Delta_{if}  \Biggl(   [A^2]_{if} \bigl(  (k_i-1)(k_f-1) +1-  [A^2]_{if} \bigr) 
-\bigl( [A^3]_{if} \bigr)^2  +\sum_j [A^2]_{ij} [A^2]_{jf} \bigl(  a_{ij}+ a_{jf}\bigr)    \nonumber \\
 +&  \sum_{j} a_{ij}a_{jf}\Bigl( \bigl( ([A^2]_{ij})^2 + ([A^2]_{fj} )^2 \bigr)  
-\bigl( [A^2]_{ij} + [A^2]_{fj}\bigr) \Bigr) \Biggr) \nonumber \\
+ & a_{if}  \Biggl(  [A^3]_{ff} \bigl(2k_f+k_i-5\bigr)  
+[A^3]_{ii}  \bigl(2k_i+k_f-5\bigr) +[A^2]_{if} \bigl(11-3k_i-3k_f\bigr)  \nonumber \\
& -2 \sum_{j} a_{ij}a_{jf}\Bigl( \bigl( [A^2]_{ij} + [A^2]_{fj}\bigr) \Bigr)  \Biggr), \nonumber \\
R^6[1,2]_{if} +&R^6[1,3]_{if} = -\Delta_{if} \Biggl(  [A^2]_{if} \bigl( [A^4]_{ii}+ [A^4]_{ff} \bigr) 
+4[AGA]_{if} -\{ A^2, G^2-3G \}_{if} -\{ AGA, A^2 \}_{if} \nonumber \\ 
-4&  [A^2]_{if}  -\sum_{j}  a_{ij} a_{jf}  \biggl( \Bigl([A^2]_{if})^2+  ([A^2]_{if})^2\Bigr) 
+2\bigl( [A^3]_{ij} + [A^3]_{fj}\bigr) -\bigl( [A^2]_{ij} + [A^2]_{fj}\bigr)          \biggr)  \Biggr)  \nonumber \\
+&a_{if} \Biggl( -2[A^2]_{if} [A^3]_{if} +2[A^2]_{if}(k_i+k_f-3) 
+2 \sum_{j} a_{ij}a_{jf} \bigl( [A^2]_{ij} + [A^2]_{fj}\bigr) \Biggr),
\nonumber \\
R^6[1,4]_{if} =&   [A^3]_{if}\Delta_{if} \Bigl(  ([A^3]_{if})^2 - 3  [A^2]_{if} +2    \Bigr). 
\end{align}

$R^6[2]_{if} $, $R^6[3]_{if} $ and $R^6[4]_{if}  $ are respectively given by the following expressions;
\begin{align}
R^6[2]_{if} +&R^6[3]_{if} = a_{if} \Biggl(  2[A^4]_{if} -( \bigl( [A^5]_{ii} + [A^5]_{ff}\bigr)  
-7\bigl( [A^3]_{ii} + [A^3]_{ff}\bigr)  +22[A^2]_{ij}  \nonumber \\
& +4[A^3]_{if}[A^2]_{if} +2\bigl( [A^3]_{ii}k_i + [A^3]_{ff}k_f\bigr)+\sum_{j} [A^3]_{jj} \bigl(a_{jf}+a_{ij}\bigr)  
\nonumber \\
 &-4\sum_{j} a_{ij}a_{jf} \bigl( [A^2]_{ij}    + [A^2]_{fj}\bigr)  -6\{A^2,G \}_{if} -2[AGA]_{if} +\{A,AGA\}_{ii}+\{A,AGA\}_{ff}  \Biggr),  \nonumber \\
R^6[4]_{if} =&  a_{if} \Biggl( [A^4]_{if} - [AGA]_{if} -\{A^2,G\} +5 [A^2]_{if}  - \Bigl( [A^3]_{ii} +  [A^3]_{ff} \Bigr) \Biggr). 
\end{align}

By unifying all the terms, we obtain the full expression of $R^6$. 
It is too long and complex that we do not describe it here.   
Lastly we give the expressions of Tr $R^n$ appearing in Eq. (7).  

 \begin{align}
Tr (R^2) &=0,   \nonumber  \\
Tr (R^3)  &=Tr (A^3),  \nonumber  \\
Tr(R^4)   &= Tr(A^4)-3 Tr(GA^2), +2Tr(A^2) + Tr(G^2-G), \nonumber \\
Tr(R^5)   &= Tr(A^5)-3 Tr(GA^3) +6 Tr(A^3) -diag(A^3) Tr(A^2)  +Ndiag(2A^3G-A^3),  \nonumber \\
Tr(R^6)&= Tr(A^6) +6Tr(A^4)-5Tr(GA^4) -4Tr(A^3)  +Tr(A^2G^2)  -6Tr(A^2G)+4Tr(A^2) \nonumber \\
&+2Tr(AGAG)   -\sum_i (a_{ii})^2 -\sum_{i,j} [A^3]_{jj}a_{ij}[A^2]_{ij} + 6  \sum_{i,j} a_{ij}[A^2]_{ij} +\sum_{i,j,k} a_{ij} a_{jk}  a_{ki}  [A^2]_{jk}.  
\end{align}

\end{document}